\documentclass[aip,amsmath,amssymb,reprint,twocolumn]{revtex4-1}
\usepackage{graphicx}
\usepackage{dcolumn}
\usepackage{bm}
\usepackage[utf8]{inputenc}
\usepackage[T1]{fontenc}
\usepackage{mathptmx}

\draft 

\begin{document}

\title{Accurate phonon blockade detector composed of a quadratically coupled optomechanical system} 



\author{Ye-Xiong Zeng}
\affiliation{School of Physics, Dalian University of Technology,  Dalian, 116024, China}
\author{Tesfay Gebremariam}
\affiliation{Department of Physics, College of  Natural Sciences, Arba Minch University, P. O. Box 21, Arba Minch, Ethiopia}
\author{Jian Shen}
\affiliation{Fundamental Education College, Dalian Neusoft University of Information, Dalian 116023, China}
\author{Biao Xiong}
\affiliation{College of Physics and Electronic Science, Hubei Normal University, Huangshi 435002, People’s
Republic of China}
\author{Chong Li}
\email{lichong@dlut.edu.cn}
\affiliation{School of Physics, Dalian University of Technology,  Dalian, 116024, China}


\date{\today}
\begin{abstract}
The observation of phonon blockade in a nanomechanical oscillator is clear evidence of its quantum nature. However, it is still a severe challenge to measure the strong phonon blockade in an optomechanical system with effective nonlinear coupling.  In this paper, we propose a theoretical proposal for detecting the phonon blockade effect in a quadratically coupled optomechanical system by exploiting supervised machine learning. The detected optical signals are injected into the neural network as the input, while the output is the mechanical equal-time second-order correlation. Our results show our scheme performs superior performance on detecting phonon blockade. Specifically, it is efficient for nonlinear coupling systems; it performs a high precision for strong photon blockade; it is robust against the small disturbance of system parameters.  Our work opens a promising way to build a phonon blockade detector.
\end{abstract}

\pacs{}

\maketitle 

Optomechanical systems, describing resonant optical cavities or electrical circuits coupled with the macroscopic mechanical structure via radiation pressure, have attracted continuous attention and made tremendous advances~\cite{RevModPhys.86.1391}.  Specifically, it provides an ideal platform for macroscopic quantum experiments, for example preparing entangled states~\cite{Wang2013,Vitali2007,Tian2013,Palomaki2013}, cooling mechanical oscillators~\cite{OConnell2010,Riedinger2018,PhysRevLett.92.075507,PhysRevLett.110.153606,PhysRevA.92.033841,PhysRevA.89.053821,Chan2011,PhysRevB.80.144508,PhysRevB.78.134301,Dong2012Optomechanical}, designing single-photon sources~\cite{PhysRevLett.107.063601,PhysRevLett.121.153601,PhysRevLett.99.073601,PhysRevA.98.023856,PhysRevA.92.023838,PhysRevA.92.033806,PhysRevA.98.043858,PhysRevA.92.053837,PhysRevA.98.033825,PhysRevA.90.033809,PhysRevA.90.043822}, realizing quantum synchronization~\cite{PhysRevLett.111.103605,PhysRevA.101.013802,PhysRevLett.123.017402}, and so on~\cite{doi:10.1063/1.4889804,doi:10.1063/1.2711181,doi:10.1063/1.3513213,doi:10.1063/1.4709416}. Besides, the theoretical predictions of many physical phenomena in quadratically coupled optomechanical systems  are consistent with the measurement in the experiment~\cite{Lee2015Multimode,Jayich2010Strong,PhysRevX.1.021011}. All of these significant efforts have enhanced the development of quantum theory as well as quantum information science.

More recently, considerable attention focus on generating and manipulating a few phonons in optomechanical systems~\cite{guan2017phonon}. In particular, phonon blockade, as a pure quantum effect, describing mechanical oscillators limited to low phonon level, has been an important research topic in macroscopic quantum word~\cite{PhysRevA.93.033846,Pogosov2008}. A lot of theoretical and experimental protocols have discussed how to achieve a phonon blockade in various quantum systems. These protocols mainly origin from two physical mechanisms. The first is to design an appropriate nonlinear Hamiltonian so that the level structure of the system is modified and the external driving is decoupled with the system~\cite{PhysRevA.100.063840}. The second is to achieve the destructive interference effect between different two jump paths and then the higher phonon excitation is restrained, which is the so-called unconventional phonon blockade~\cite{wang2018unconventional}. To summarize, these theoretical and experimental efforts provide a cornerstone for the detection of phonon blockade and quantum information processing. 
However, some severe challenges are obstructing the detection of phonon blockade in the current experiment platform~\cite{PhysRevA.82.032101,PhysRevA.93.013808,PhysRevA.94.063853,PhysRevB.84.054503,PhysRevA.98.013821,PhysRevA.99.013804,PhysRevA.93.063861,Shi2018}. To overcome these difficulties, Didier et al. have proposed an effective detection scheme for phonon blockade by utilizing a superconducting microwave resonator linearly coupled to the mechanical oscillator for transducing its motion into an electric signal~\cite{PhysRevB.84.054503}. Though this protocol is effective and feasible in hybrid superconducting systems with linear interactions, it is infeasible in mechanical systems with nonlinear interactions. In particular, the nonlinear coupling is an intrinsic character of  optomechanical systems~\cite{book1429473}. In theory and experiment, people achieve the phonon blockade by utilizing an unequal interval energy level or quantum interference effect, and thus one usually has to create nonlinear interactions~\cite{Sarma2018,Hai2019Phonon,PhysRevA.96.013861}. Besides, Cohen et. al. declared have observed phonon blockade in a linearized optomechanical system by measuring the optical field~\cite{Cohen2015}. This protocol is efficient for measuring the phonon bunching while it is impracticable to detect the strong phonon antibunching due to the tiny motion of a mechanical oscillator closing to its ground state. Therefore, it is significant and imperative to propose a more appropriate detecting scheme that can be applied to detect the strong antibunching effect in the mechanical systems with nonlinear interactions.

Machine learning provides an opportunity to acquire the nonlinear function map between the input and the output of a neural network~\cite{PhysRevE.98.053304,PhysRevE.99.023304,PhysRevE.99.062701}. It is based on the universal approximation theorem~\cite{Hornik1991,Rev045002} that continuous functions on compact subsets $\mathfrak{R}$ can be approximated by a feedforward network composed of a single hidden layer and a finite number of neurons under mild assumptions of the activation function. Therefore, the feedforward network may have a potential application in detecting the phonon blockade effect.  Furthermore, machine learning techniques have been widely applied to study quantum physics, including controlling quantum system~\cite{Hou2019,PhysRevX.8.031086,Niu2019,Chen2014}, cooling mechanical oscillator~\cite{sommer2019prospects},  transporting quantum coherent state~\cite{Porotti2019}, reconstructing quantum dynamics~\cite{PhysRevX.10.011006} and many others. In particular, some previous efforts have trained some neural networks to predict other significant theoretical and experimental results~\cite{Ghosh2019,PhysRevA.97.042324}. Therefore, machine learning techniques may be an ideal tool to detect the phonon blockade in a nonlinear optomechanical system.

Inspired by these works, we propose a supervised machine learning approach towards contriving a strategy for detecting phonon blockade.
We consider a quadratically coupled optomechanical system with a membrane in the middle of the configuration. Specifically, we obtain an effective $\chi^{(2)}$ nonlinear coupling~\cite{Li:18,PhysRevA.90.023849,Xiong:20,Zeng:20} in the quadratically coupled optomechanical system by coherently driving the cavity and the mechanical oscillator. 
The strong phonon blockade appears in this nonlinear system by adjusting the parameters of the system. Moreover, we consider the detected results of the cavity mode as the input of a feedforward neural network and the mechanical equal-time second correlation as the output. The numerical results exhibit that the trained neural network can estimate the equal-time second-order correlation function with high precision even a strong phonon blockade appears in the optomechanical system with nonlinear interaction. Besides, our scheme has strong robustness for the instability of the system parameters. In particular, this scheme can be applied to build a phonon blockade detector.

\begin{figure}[ht]
\centering 
\includegraphics[width=8cm]{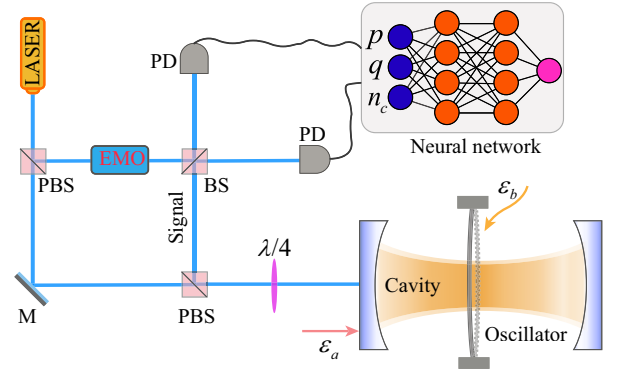}
\caption{Schematic diagram of detecting phonon blockade effect by machine learning in a quadratically coupled optomechanical system. The monochromatic cavity field is coupled to a thin movable mechanical oscillator. The laser splits into a signal beam and a local oscillator (LO). The cavity is driven by the signal beam. The LO is phase modulated via an electro-optical modulator (EOM) and utilized to achieve homodyne detection together with the output of the optical cavity. The detected optical signals are acted as the input of a neural network. Meanwhile, the cavity field and the mechanical oscillator are driven by two weak driving lasers $\varepsilon_a$ and $\varepsilon_b$, respectively.}
\label{fig1}
\end{figure}

Specifically, we propose a quadratically coupled optomechanical system which consists of an optical cavity and a mechanical oscillator, where the optic cavity is quadratically coupled to the mechanical oscillator, as schematically illustrated in Fig.~\ref{fig1}. The quadratically nonlinear interaction between the cavity field and mechanical oscillator is obtained by attaching the thin mechanical oscillator at the node (or antinode) of the intracavity standing-wave existed in a Fabry-P\'{e}rot cavity. Besides, we assume the optical cavity is driven by a strong driving field.  Meanwhile, the mechanical and optical modes are driven by two thin classical lasers with strengths ($\varepsilon_a$, $\varepsilon_b$) and frequencies ($\omega_a$, $\omega_b$), respectively. For this model, under the frame rotating with the driving frequency $\omega_L$, the Hamiltonian of the system is written as  ($\hbar=1$)
\begin{equation}
\begin{aligned} 
\hat{H}=& \Delta_{c} \hat{a}^{\dagger} \hat{a}+\omega_{m} \hat{b}^{\dagger}\hat{b}+g \hat{a}^{\dagger} \hat{a}\left(\hat{b}^{\dagger}+\hat{b}\right)^{2} \\
 &+\left(\Omega_{L} \hat{a}^{\dagger}+\varepsilon_{a} e^{-i \delta_{a} t} \hat{a}^{\dagger}+\varepsilon_{b} e^{-i \omega_{b} t} \hat{b}^{\dagger}+\mathrm{H.c.}\right), \end{aligned}
\label{eq1}
\end{equation}
where $\hat{a}$ ($\hat{a}^{\dagger}$) and $\hat{b}$ ($\hat{b}^{\dagger}$) are the annihilation (creation) operator of the single-mode cavity field and the mechanical oscillator with the respective resonant frequencies $\omega_c$ ($\Delta_c=\omega_c-\omega_L$ ) and $\omega_m$. $\vert\Omega_L\vert=\sqrt{2P\kappa/\omega_L}$ represents the strong driving magnitude where $P$ describes the input power of the strong driving field and $\kappa$ represents the decay of the cavity mode. The detuning between the two driving fields of the cavity is $\delta_a=\omega_a-\omega_L$. The quadratic optomechanical interaction between the cavity field and the mechanical oscillator is described by the third term in Eq.~(\ref{eq1}) and the corresponding coupling coefficient is $g$. 

\begin{figure}[ht]
\centering 
\includegraphics[width=8cm]{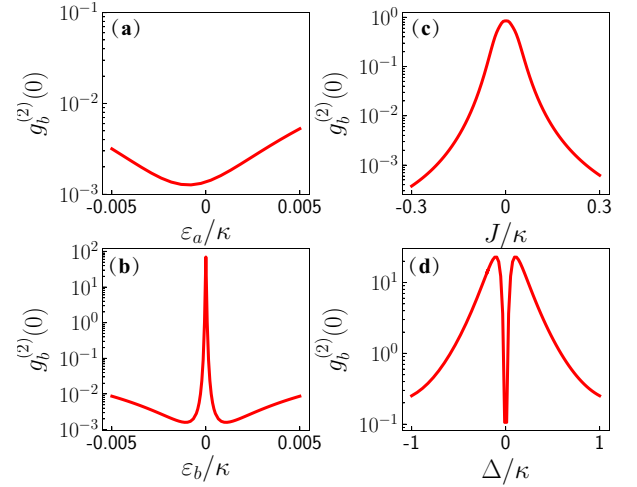}
\caption{(a) The equal-time second-order correlation function $g^{(2)}(0)$ versus the optical driving strength $\varepsilon_a$ with $\Delta=\Delta_a=\Delta_b=0$, $J/\kappa=0.2$, and $\varepsilon_b/\kappa=0.002$.  (b) $g^{(2)}(0)$ is plotted as a function of the mechanical driving $\varepsilon_b$ with $\Delta=\Delta_a=\Delta_b=0$, $J/\kappa=0.2$, and $\varepsilon_a/\kappa=0.002$. 
 (c) Plot of $g^{(2)}(0)$ versus the coupling strength $J$ under $\Delta=\Delta_a=\Delta_b=0$ and $\varepsilon_a=\varepsilon_b=0.002$. (d) Plot of $g^{(2)}(0)$ versus the detune $\Delta$ where $\Delta=\Delta_a=\Delta_b$, $\varepsilon_a/\kappa=\varepsilon_b/\kappa=0.002$, and $J/\kappa=0.2$. 
Other parameters are $\gamma/\kappa=0.001515$ and $n_{\text{th}}=10^{-3}$.}
\label{fig2}
\end{figure} 
We suppose that the classical laser driving amplitude $\Omega_L$ is sufficiently large, i.e., $\Omega_L\gg\kappa,\gamma$ and the two thin driving fields $\varepsilon_a$, $\varepsilon_b$ are far less than the decay of the cavity, that is, $\varepsilon_a,\varepsilon_b\ll\kappa$. Taking these conditions into consideration, we  can replace  the operators $\hat{a}$ and $\hat{b}$ as the sum of their steady mean values and quantum fluctuation operators, that is, $\hat{a}\longrightarrow\alpha+\hat{a}$ and $\hat{b}\longrightarrow\beta+\hat{b}$, where the steady-steady mean value is written as 
\begin{subequations}
\begin{equation}
\alpha=\frac{2 \Omega_{L}}{-2\Delta_{c}+i\kappa},
\label{eq2a}
\end{equation}
\begin{equation}
\beta=0,
\label{eq2b}
\end{equation}
\end{subequations}
by approximatively ignoring the small contribution of the two thin driving fields $\varepsilon_a$ and $\varepsilon_b$. The optical mean value $\alpha$ is much larger than the quadratic coupling strength $g$ and the fluctuation $\hat{a}$ so that the $g\hat{a}^{\dagger}\hat{a}\left(\hat{b}^{\dagger}+\hat{b}\right)^{2}$ can be neglected. Then we can obtain the following Hamiltonian
\begin{equation}
\begin{split}
\hat{H}^{\prime}=& \Delta_{c} \hat{a}^{\dagger} \hat{a}+\omega_{m} \hat{b}^{\dagger} \hat{b}+g\left(|\alpha|^{2}+\hat{a}^{\dagger} \hat{a}\right)\left(\hat{b}^{\dagger}+\hat{b}\right)^{2} \\
&+g\left(\alpha \hat{a}^{\dagger}+\alpha^{*} \hat{a}\right)\left(\hat{b}^{\dagger}+\hat{b}\right)^{2} \\
&+\left(\varepsilon_{a} e^{-i \delta_{a} t} \hat{a}^{\dagger}+\varepsilon_{b} e^{-i \omega_{b} t} \hat{b}^{\dagger}+\mathrm{H} \mathrm{c} .\right).
\end{split}
\label{eq3}
\end{equation}
For convenience of computation, we study the dynamics of the system under the rotating reference frame with respect to the unitary operator $U(t)=\exp \lbrace i \delta_{a}\hat{a}^{\dagger}\hat{a}+i \omega_{b}\hat{b}^{\dagger}\hat{b}\rbrace$. Therefore, by considering the rotating-wave approximation, i.e., neglecting the fast oscillation term, the effective Hamiltonian $\hat{H}_{\mathrm{eff}}=\hat{U}\hat{H}'\hat{U}^{\dagger}-i\hat{U}d\hat{U}^{\dagger}/dt$ can be simplified as 
\begin{equation}
\begin{aligned} \hat{H}_{\mathrm{eff}}^{\prime}=& \Delta_{a} \hat{a}^{\dagger} \hat{a}+\Delta_{b}\hat{b}^{\dagger} \hat{b}+\left(J\hat{a}\hat{b}^{\dagger 2}+J^*\hat{a}^{\dagger} \hat{b}^{2}\right) \\ &+\left(\varepsilon_{a} \hat{a}^{\dagger}+\varepsilon_{b} \hat{b}^{\dagger}+\mathrm{H.c.}\right), \end{aligned}
\label{eq4}
\end{equation}
where the effective parameters are $J=g\alpha$, $\Delta_a=\Delta_c-\delta_a$  and $\Delta_{b}=\omega_{m}+2 g|\alpha|^{2}+g-\omega_{b}$. Moreover, we have postulated the frequencies and detunes meeting the following conditions $\delta_a=2\omega_d$ and $\Delta_a, \Delta_b\ll\omega_b,\omega_m$. 

In order to  both theoretically and experimentally measure the phonon blockade effect, researchers usually adopt the equal-time second-order correlation function of the steady state, i.e.,
\begin{equation}
g_{b}^{(2)}(0) = \frac{\left\langle b^{\dagger}b^{\dagger}bb\right\rangle}{\langle\hat{b}^{\dagger}\hat{b}\rangle^{2}}=\frac{\operatorname{Tr}\lbrace\hat{b}^{\dagger}\hat{b}^{\dagger}\hat{b}\hat{b}\hat{\rho}_s\rbrace}{\operatorname{Tr}\lbrace\hat{b}^{\dagger}\hat{b}\hat{\rho}_s\rbrace^2}.
\label{eq5}
\end{equation}
To calculate the correlation function under the phonon or photon blockade, people usually expand the state of the system to the Fock-state space whose basis is $|mn\rangle$, where $m$ denotes the energy level of  the optical cavity and $n$ represents the energy level of  the mechanical oscillator~\cite{PhysRevA.90.033809,Ferretti_2013}.  In these works, they truncate the system to the few photon or phonon space spanned by $\vert 00\rangle$, $\vert 10\rangle$, $\vert 01\rangle$, and $\vert 02\rangle$ because excitations of high energy level can be negligible.

For our scheme,  we consider wider parameter ranges to detect the phonon blockade effect rather than confine a group of parameters for appearing phonon blockade. Therefore, it is necessary to truncate the system to a higher space~\cite{JOHANSSON20131234,JOHANSSON20121760}. 
 \begin{figure}[ht]
\centering 
\includegraphics[width=8cm]{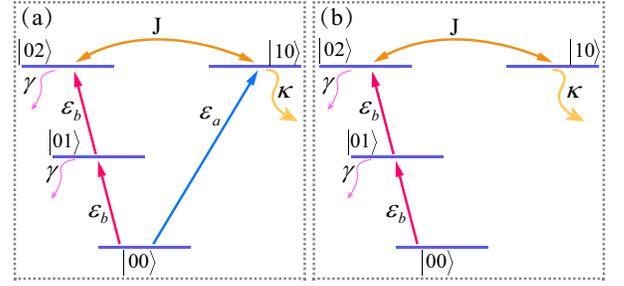}
\caption{(a) Energy-level diagram. The zero-, single-, and two-phonon states (horizontal blue short lines) and the transition paths leading to the quantum interference responsible for the strong antibunching (the red and the blue lines with arrows). (a) and (b) are for different driving. (a) Bimodal driving. (b) The weak optical driving is zero, i.e., $\varepsilon_a=0$.}
\label{fig3}
\end{figure}
To this aim , we adopt the Lindblad markovian master equation to govern the dynamics of the system, which allows us to numerically calculate a higher energy level. The master equation  has the following expression
\begin{equation}
\begin{split}
\frac{d\hat{\rho}}{d t}=&-i\left[\hat{H}^{\prime}_{\mathrm{eff}}, \hat{\rho}\right]+\frac{\kappa}{2}\left(2 \hat{a} \hat{\rho} \hat{a}^{\dagger}-\hat{a}^{\dagger} \hat{a} \hat{\rho}-\hat{\rho} \hat{a}^{\dagger} \hat{a}\right) 
\\ 
&+\frac{\gamma}{2}(n_{\mathrm{th}}+1)\left(2 \hat{b}\hat{\rho}\hat{b}^{\dagger}-\hat{b}^{\dagger} \hat{b} \hat{\rho}-\hat{\rho} \hat{b}^{\dagger} \hat{b}\right) 
\\
 &+\frac{\gamma}{2} n_{\mathrm{th}}\left(2\hat{b}^{\dagger}\hat{\rho} \hat{b}- \hat{b}\hat{b}^{\dagger}\hat{\rho}-\hat{\rho} \hat{b} \hat{b}^{\dagger}\right), 
\end{split}
\label{eq6}
\end{equation}
where we have assumed that the cavity is placed in a vacuum bath while the oscillator  is situated in a heat environment with temperature $T_m$. Accordingly,  the average thermal phonon numbers of the mechanical oscillator is $n_{\mathrm{th}}=(\exp(\omega_m/k_BT_m)-1)^{-1}$, where $k_b$ is the the Boltzmann constant. In experiment, the mechanical oscillator can reach a low temperature $T=25~\text{mK}$ with frequency 6~GHz, and thus  the average thermal phonon numbers $n_{\text{th}}$ can arrive a small value about $10^{-5}$ by designing a conventional dilution refrigerator~\cite{OConnell2010}. Therefore, it is reasonable to assume a small average thermal phonon numbers $n_{\text{th}}=10^{-3}$ for our scheme.

According to Eq.~(\ref{eq6}), we numerically simulate the equal-time second-order correlation function $g^{(2)}_b(0)$ as a function of the weak optical driving strength $\varepsilon_a$ in Fig.~\ref{fig2}(a). It is obvious that the correlation function $g^{(2)}_b(0)$ change slowly with the varying of the driving strength $\varepsilon_a$. Moreover, the correlation $g^{(2)}_b(0)\ll 1$ exhibits a strong phonon blockade in a small range of driving $\varepsilon_a$. In Fig.~\ref{fig2}(b), we show the dependence of the correlation function $g^{(2)}_b(0)$ on the mechanical driving strength $\varepsilon_b$. We noted that there is a narrow peak for $g^{(2)}_b(0)>1$ near to $\varepsilon_b=0$ and the small $g^{(2)}_b(0)$ are obtained in the other wide region, i.e., $g^{(2)}_b(0)\ll 1$. The correlation function $g^{(2)}_b(0)$ is drawn as the function of the effective coupling coefficient $J$ in Fig.~\ref{fig2}(c). The corresponding results show that $g^{(2)}_b(0)$ decreases with the increase of $\vert J\vert$ and the blockade appears in whole achievable range of the coupling strength $J$.  In order to be more rigorous, we examine the variation tendency of $g^{(2)}_b(0)$ with the changing of the detuning $\Delta=\Delta_a=\Delta_b$.  From fig.~\ref{fig2}(d), one can observe the most strong blockade effect appeared at $\Delta=0$. However, a small disturbance of $\Delta$ will lead to a tremendous modification of $g^{(2)}_b(0)$.  

On one hand, the physics behind the phonon antibunching is mainly due to the quantum interference effect existed between two different pathways, as shown in Fig.~\ref{fig3}(a). There are two transition paths from single-phonon state to two-phonon state. One is the direct path pumped by the weak classical laser, i.e., $|00\rangle \stackrel{\varepsilon_b}{\longrightarrow}|01\rangle \stackrel{\varepsilon_b}{\longrightarrow}|02\rangle$, and the other path is through mode optical mode $\hat{a}$. Exactly, the hybrid optomechanical system starts from the ground state $\vert 00\rangle$. Then the system is excited to the single-photon state by the weak optical driving. At last, the system arrives at the two-phonon state via the second-order nonlinear interaction between the mechanical and optical modes. The whole path can be expressed in mathematical symbols   
, that is, $|00\rangle \stackrel{\varepsilon_{a}}{\longrightarrow}|10\rangle \stackrel{J}{\longrightarrow}|02\rangle$. The excitons coming from the two pathways will  perform the destructive interference phenomenon.
 One the other hand, as shown in Fig.~\ref{fig3}(b), both the decay $\kappa$ of the optical cavity and the coupling strength $J$  are far greater than $\gamma$, $\varepsilon_a$, and $\varepsilon_b$. The whole system is driven by the mechanical driving.  Under these conditions, the energy of the system automatically decays to the bath via the decay channel of the optical cavity, 
i.e., $|00\rangle \stackrel{\varepsilon_{b}}{\longrightarrow}|01\rangle
\stackrel{\varepsilon_{b}}{\longrightarrow}|02\rangle
 \stackrel{J}{\longrightarrow}|10\rangle\stackrel{\kappa}{\longrightarrow}\text{environment}$, which can also promote the whole system steadying in low excitation space. References \cite{PhysRevA.92.023838,Shi2018} provided similar results as Figs.~\ref{fig2} and~\ref{fig3}and these results will guide us to obtain the sample data, but their aim is to find the ranges of systemic parameters existing phonon blockade effect and they leave a severe challenge: how to detect the phonon blockade effect. In our work, we aim to address these remaining challenges.

In this section, we introduce our scheme in detail. For the convenience of description, we start from the Langevin equation 
\begin{equation}
\begin{split}
&\frac{d\hat{a}}{dt}=(-i\Delta_a-\frac{\kappa}{2})\hat{a}-iJ^*\hat{b}^{2}-i\varepsilon_a+\sqrt{\kappa}\hat{a}_{in}, \\
&\frac{d\hat{b}}{dt}=(-i\Delta_b-\frac{\gamma}{2})\hat{b}-2iJ\hat{a}\hat{b}^{\dagger}-i\varepsilon_b+\sqrt{\gamma}\hat{b}_{in},
\end{split}
\label{eq7}
\end{equation}
where $\hat{a}_{in}$ and $\hat{b}_{in}$ represent the input noise of the surrounding environment and satisfy the two-time correlation function $\langle\hat{a}_{in}(t)\hat{a}^{\dagger}_{in}(t^{\prime})\rangle=\delta(t-t^{\prime})$ and $\langle\hat{b}_{in}(t)\hat{b}^{\dagger}_{in}(t^{\prime})\rangle=(n_{\text{th}}+1)\delta(t-t^{\prime})$. Besides, the output of cavity meets the following relation     
\begin{equation}
\hat{a}_{out}=\sqrt{\kappa}\hat{a}-\hat{a}_{in}.
\end{equation}
\label{eq8}
We note that the Eq.~(\ref{eq7}) is nonlinear Langevin equations, which means the phonon correlation function is hard to be represented by the output of the optical signal. Moreover,  it is hard to directly measure the phonon mode by using the current experimental technique. Therefore,  almost all of the experiments measuring the mechanical oscillator are indirectly measuring the output of the optical field. As we mentioned before, one usually transfers the mechanical motion to optical or electrical signals in the experiment when the mechanical oscillator is linearly coupled to the optical mode or electrical mode. However, Eq.~(\ref{eq7})  indicates that there is a complex nonlinear relation existed in the optical field and the mechanical oscillator. Moreover, when a strong phonon blockade is obtained, the motion of the mechanical oscillator is thin due to close to the ground state. Therefore, the difficulty of the detection is largely increased. Therefore, the primary challenge is how to find the nonlinear relation between the mechanical and optical or electrical signals and measure the weak mechanical signal. Supervised machine learning is famous for its ability in dealing with nonlinear problems. Therefore, we discuss how to detect the phonon blockade effect by applying the supervised machine learning technique.

The specific realization is depicted in Fig.~\ref{fig1}. The output of the optical mode is detected by homodyne detection and the corresponding measurement results are considered as the input of a trained feedforward neural network. The feedforward neural network is composed of an input layer, some pattern layers, and an output layer. In feedforward neural networks, the signal flows from the input layer to the output layer without feedback loops. A multilayer perceptron is obtained when a feedforward neural network has one or more hidden layers. With enough neurons, a multilayer perceptron can approximate any continuous nonlinear function and solve many complicated tasks. Here, we consider a network model where the hidden layer has 50 nodes, i.e., $L=50$. We have optimized the numbers of the nodes and the hidden layer optimal. The complexity of the neural network is suitable and enough to make us obtaining good precision. The input layer has three nodes which in our scheme corresponds to the mean photon number $n_c=\langle\hat{a}^{\dagger}\hat{a}\rangle$, the optical quadratures $p=\langle \hat{p}\rangle$ and $q=\langle\hat{q}\rangle$ of steady state, where we define the operators $\hat{q}=\frac{\hat{a}+\hat{a}^{\dagger}}{\sqrt{2}}$ and $\hat{p}=\frac{\hat{a}-\hat{a}^{\dagger}}{\sqrt{2}i}$.

Besides, we consider a single node as the output of the neural network and the output node gives the final results, i.e., $\log_{10}\left(g^{(2)}_b(0)\right)$. Each artificial neuron is described by a real activation function $\phi(\mathbf{X},\mathbf{w},\mathbf{b})$ parametrized by a real matrix of weights $\mathbf{w}$ and a vector of bias $\mathbf{b}$, where $\mathbf{X}$ is the input vector. The training task  is to optimize the weight matrix $\mathbf{w}$ and the bias vector $\mathbf{b}$, and then minimize the error in the training process. Here we adopt the tanh-function $\phi(x)=\tanh(x)$ as the activation function. The neural network transfers the information between two layers by the feedforward method where  weights and bias are optimized via Levenberg-Marquardt algorithm~\cite{gardiner2004quantum}.

The whole sample set is defined as
\begin{equation}
S=\lbrace \left(\text{X}^{1},\text{Y}^{1}\right), \left(\text{X}^{2},\text{Y}^{2}\right),\cdots, \left(\text{X}^{N},\text{Y}^{N}\right)\rbrace.
\label{eq9}
\end{equation} 
where $N$ is the number of the samples. In the $k$th sample, the input vector is $\text{X}^{(k)}=\lbrace p^{(k)}, q^{(k)}, n^{(k)}_c\rbrace$ and the corresponding output  is $\text{Y}^{(k)}=\lbrace \log_{10}\left(g^{(2)}_b(0)\right)\rbrace_k$. The sample set consists of a train set $S_{\text{train}}$, a test set $S_{\text{test}}$, and a validation set $S_{\text{val}}$, i.e., $S=S_{\text{train}}\cup S_{\text{test}}\cup S_{\text{val}}$. The training set is utilized to estimate the model while the verification set is applied to determine the parameters of the network structure or the complexity of the control model. The test set is applied to test the performance of the final selection of the optimal model. Moreover, the performance of the neural network  could be tested by the mean squared error (MSE) between the ideal $Y^k$ of any training (or testing or validation) and the estimated output $Y^{\prime k}$ from by neural network. The MSE is calculated via the following expression
\begin{equation}
\textit{MSE}=\frac{1}{M}\sum^{M}_{k=1}(Y^k-Y^{\prime k})^T(Y^k-Y^{\prime k}),
\label{eq10}
\end{equation}
where $M$ is the number of training (or testing or validation) samples. The iteration number of the neural network can be determined by checking the mean squared error (or the classification success rate) for the testing set. Finally, the trained neural network will be used to detect the equal-time second-order correlation function for new input vectors (out of the training set), where the input of the neural work is obtained by the homodyne detection.

\begin{figure}[ht]
\centering 
\includegraphics[width=8cm]{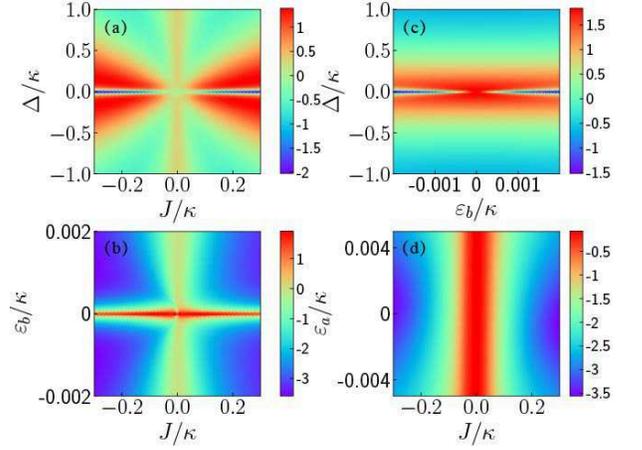}
\caption{(a) The equal-time second-order correlation function $g^{(2)}(0)$ versus coupling strength $J/\kappa$ and  detune $\Delta/\kappa$~($\Delta=\Delta_a=\Delta_b$) with $\varepsilon_a/\kappa=\varepsilon_b/\kappa=0.002$. (b) $g^{(2)}(0)$ is plotted as a function of the coupling strength $J/\kappa$ and the mechanical driving $\varepsilon_b/\kappa$ with $\Delta=\Delta_a=\Delta_b=0$ and $\varepsilon_a/\kappa=0.002$. 
 (c) Plot of $g^{(2)}(0)$ versus the detune $\Delta/\kappa$ ($\Delta=\Delta_a=\Delta_b$) and the mechanical driving $\varepsilon_b$ under the coupling strength $J/\kappa=0.2$ and $\varepsilon_a/\kappa=0.002$. (d) Plot of $g^{(2)}(0)$ versus the coupling $J/\kappa$ and the optical driving $\varepsilon_a/\kappa$ where $\Delta=\Delta_a=\Delta_b=0$ and $\varepsilon_b/\kappa=0.002$. Other parameters are $\gamma/\kappa=0.001515$ and $n_{\text{th}}=10^{-3}$.}
\label{fig4}
\end{figure}

Generally, the different sample set is required for various problems. For our scheme, it is better to traverse wide the parameter range, but the sample set is too large to be calculated. Moreover, it may reduce the training speed. To speed up the training process and obtain an effective trained neural network, we have to find an appropriate sample set with the finite sample. To handle this knotty problem we focus on the effective parameters of the system. The effective system Hamiltonian $H^{\prime}_{\text{eff}}$  includes four effective parameters $J$, $\Delta=\Delta_a=\Delta_b$, $\varepsilon_a$, $\varepsilon_b$, which will affect the detect result. Therefore, we simulate the equal-time second-order function of the steady state as the function of any two different parameters among $J$, $\Delta$, $\varepsilon_a$, $\varepsilon_b$ and the corresponding numerical results are given in Figs.~\ref{fig4}(a)-(d). These parameters are limited to a feasible range in the current experiment platform.

In Fig. \ref{fig4}, we observe that the strong phonon blockade appears in both the negative and positive areas of the parameters and the two areas are approximately comparable. In experiment, one can limit the parameters in the positive interval to obtain the strong phonon blockade effect. These reasons allows us only to study the positive range of $J,\varepsilon_a,\varepsilon_b\geq 0$. In Figs.~\ref{fig4}(a) and (c), we notice that the strong phonon blockade effect appear in a small range of detuning $\Delta/\kappa\in [-0.1, 0.1]$. Generally, the detune $\Delta$ is manipulated to the resonant point, i.e., $\Delta=0$, and then a strong blockade phenomenon appears. However, the actual detune may have a minor deviation in a small range around zero due to some disturbances existed in the system. Therefore, we can limit the region of the detune to a small interval, i.e., $\Delta/\kappa\in [-0.1, 0.1]$. 
Moreover, from Figs.~\ref{fig4}(a), (b) and (d), we observed that a more strong and stable phonon blockade effect has appeared in the around of $J/\kappa=0.3$. Thus, it is significantly important to limit the coupling coefficient $J$ in an effective and feasible region $[0.1, 0.35]$, which can largely reduce the level of difficulty and intensity of training. As shown in Fig.~\ref{fig4}(d), the correlation function $g^{(2)}_b(0)$ is not very sensitive for the weak optical driving $\varepsilon_a$ and the optimum blockade effect is approximately located at $\varepsilon_a/\kappa=0.002$.  
In Figs.~\ref{fig4}(b) and (c), the correlation function $\log_{10}\left(g^{(2)}_b(0)\right)<0$ is mainly located in a small range $\varepsilon_b/\kappa\in[0.001, 0.003]$ so that one can achieve the phonon blockade by adjusting the mechanical driving strength $\varepsilon_b$ in this region. In our scheme, we limit the mechanical driving strength $\varepsilon_b/\kappa\in[0.001, 0.003]$, which not only reduces the computation but also obtains an effective and feasible parameter range. It should be noticed that the classical driving fields are easy to fix at a stable value by stabilizing the laser intensity and thus the fluctuation is very small~\cite{PhysRevA.91.033828}.
However, the phase noise of the driving lasers usually leads to a disturbance of detuning~\cite{PhysRevA.80.063819,PhysRevA.84.032325,PhysRevA.84.063827}. Therefore, we consider a relatively larger detuning range than both driving fields to keep a high performance.  

\begin{figure}[ht]
\centering 
\includegraphics[width=8cm]{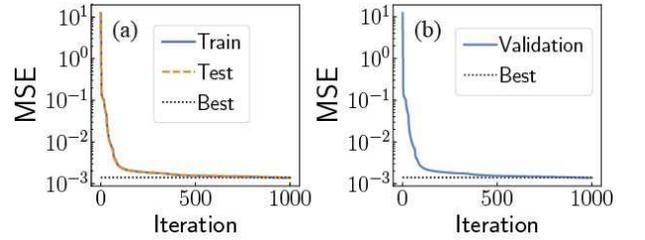}
\caption{ The training process of the feedforward neural network. (a) The mean squared error MSE for the training set and test set versus the iterations. (b)  The mean squared error MSE for validation set versus the iterations. The whole sample set includes $5\times 10^5$ examples where  $70\%$ of the whole sample set is the training set, $15\%$ is used to as the validation set and the remaining $15\%$ is considered as the test samples.}
\label{fig5}
\end{figure} 
The ultimate goal is to arrange the trained model in the actual experiment. Therefore, we want that the trained model can achieve a good prediction efficiency on the experimental data. In other words, we need to obtain the smaller the error between the prediction result and the real data. To complete this task, we have generated $5\times 10^5$ samples that are enough to overlap the small range of the system parameters. These samples are obtained by simulating $5\times 10^5$ dynamical evolution for different system parameters limited in effective and feasible regions. Moreover, to complete the supervised machine learning task, we take $3.5\times 10^5$ samples as the training set occupying $70\%$ of the whole sample set S. The $15\%$ of the whole sample set is used as the testing set to test the performance of the neural network. The remaining $15\%$ is considered as the validating set that can determine the complexity of the network structure. The training process is illustrated in Fig.~\ref{fig5}(a) and (b). On one hand, as shown in Fig.~\ref{fig5}(a), the MSEs for the training set and the testing set are plotted every six iterations. The MSEs for both training and testing sets decrease dramatically in hundreds of iterations and the output of the neural network is closer to the real value. Generally, the MSE of the training set decreases monotonously in the whole training process, while the MSE of the testing set may oscillate slightly in later iterations.  Of course, our numerical experiment shows that both MSEs of the training and testing sets decrease monotonously and can approach the minimum values of the whole training process, which means the training aim has arrived. It is because enough large sample set has been applied and the appropriate parameter range has been chosen. Therefore, it is worthwhile to obtain a superior neural network though we take a little more time to obtain more samples by simulating the system dynamics. On the other hand, we draw the MSE of the validation set in Fig.~\ref{fig5}(b). The trend of MSE of the validation set is gradually decreased to a small value, which can verify the effectiveness and feasibility of the trained neural network model. Moreover, the optimum MSEs of the training set, the testing set, and the validation set have arrived $10^{-3}$, which indicates the forecasting accuracy is close to $10^{-3}$.

\begin{figure}[ht]
\centering 
\includegraphics[width=8.2cm]{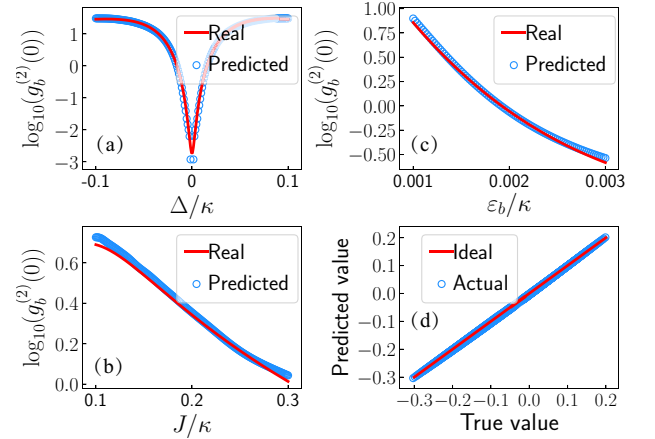}
\caption{ The numerical experiment for the trained neural network. (a) The real and predicted correlation function $\log_{10}\left(g^{(2)}_b(0)\right)$ versus the detune $\Delta/\kappa$ with $\Delta/\kappa\in [-0.1,0.1]$, $J/\kappa=0.2$ , $\varepsilon_b/\kappa=0.0015$. (b)The real and predicted correlation function $\log_{10}\left(g^{(2)}_b(0)\right)$ as the function of the effective coupling strength $J/\kappa\in [0.1, 0.3]$, $\Delta/\kappa=0.02$, $\varepsilon_b/\kappa=0.0015$. (c) The correlation function $\log_{10}\left(g^{(2)}_b(0)\right)$ as the function of the mechanical driving $\varepsilon_b/\kappa\in [0.001,0.003]$, $J/\kappa=0.2$, $\Delta/\kappa=0.02$. (d) The predicted $\log_{10}\left(g^{(2)}_b(0)\right)$ of random samples versus the real $\log_{10}\left(g^{(2)}_b(0)\right)$, where the random samples are obtained for $\Delta/\kappa\in [-0.1,0.1]$, $J/\kappa\in [0.1, 0.3]$ and $\varepsilon_b/\kappa\in [0.001,0.003]$. Other parameters are $\gamma/\kappa=0.001515$, $\varepsilon_a/\kappa=0.002$ and $n_{\text{th}}=10^{-3}$.}
\label{fig6}
\end{figure}

Here, we take some samples as examples to demonstrate the performance of the trained neural network on measuring the phonon blockade. In Figs.~\ref{fig6}~(a)-(c), we simulate the ideal as well as the detected correlation function $\log_{10}\left( g^{(2)}_b(0)\right)$ as the function of different parameters. In Fig.~\ref{fig6}(a), we demonstrate the correlation function $\log_{10}\left( g^{(2)}_b(0)\right)$  versus the detune $\Delta$ and the results show that the ideal values are approximately agreed with the detected value. Meanwhile, both the real and detected curves drawn in Figs.~\ref{fig6}~(b) and (c), are approximately coincident. Therefore, the phonon blockade can be effectively detected by inserting the measuring results of the optical field into the trained neural network,  where the given optical observable quantities are measurable with the current experimental technique.  
 To be more rigorous, we randomly choose system parameters ($\Delta$, $J$, $\varepsilon_b$) in the given range to calculate both the real and predicted values and the corresponding results are shown in ~\ref{fig6}(d). One can find that the predicted value is approximately equal to the real value. Therefore, if one has measured the mean photon number $n_c$, the optical quadratures $p$ and $q$, the equal-time second-order correlation can be efficiently estimated by the trained feedforward neural network. Moreover, our scheme is effective in a relatively large parameter range so that the trained neural network has high robustness against the disturbance of the system parameters.
 
   \begin{figure}[ht]
\centering 
\includegraphics[width=8cm]{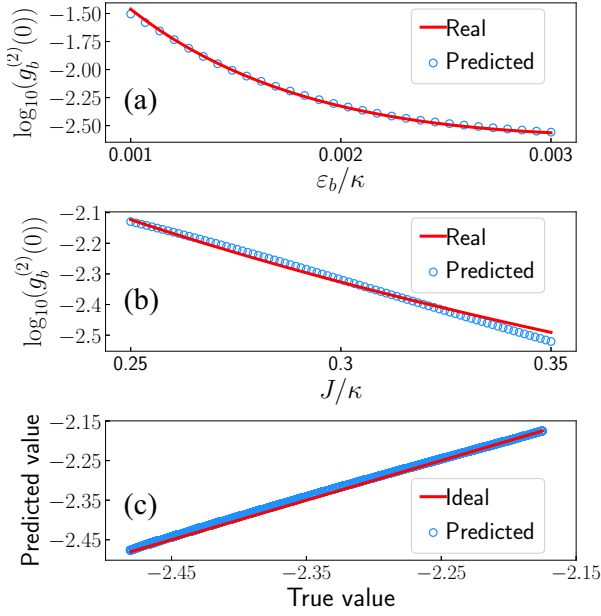}
\caption{The numerical experiment for detecting strong phonon blockade by utilizing the trained neural network. (a) The correlation function $\log_{10}\left(g^{(2)}_b(0)\right)$ as the function of the mechanical driving $\varepsilon_b/\kappa\in [0.001,0.003]$, $J/\kappa=0.3$, $\Delta/\kappa=0.005$.
(b) The real and predicted correlation function $\log_{10}\left(g^{(2)}_b(0)\right)$ as the function of the effective coupling strength $J/\kappa\in [0.25, 0.35]$, $\Delta/\kappa=0.005$, $\varepsilon_b/\kappa=0.002$. (c) The predicted $\log_{10}\left(g^{(2)}_b(0)\right)$ of random samples versus the real $\log_{10}\left(g^{(2)}_b(0)\right)$, where the random samples are obtained for $\Delta/\kappa\in [-0.005,0.005]$, $J/\kappa\in [0.29, 0.31]$ and $\varepsilon_b/\kappa\in [0.0018,0.0022]$. Other parameters are $\gamma/\kappa=0.001515$, $\varepsilon_a/\kappa=0.002$ and $n_{\text{th}}=10^{-3}$.}
\label{fig7}
\end{figure}
One of the significant and interesting points is to assess the ability of the trained neural network on detecting strong phonon blockade effect. According to the discussion above, we limit the ranges of the systemic parameters to make the mechanical oscillator obtaining a strong phonon blockade and simulate the real and predicted mechanical equal-time second-order correlation in Fig.~\ref{fig7}. One can observe that the real value is consistent with predicted value in Figs.~\ref{fig7}(a)~and~(b) when the mechanical mode appears strong phonon blockade effect.  These results verify the effectiveness of the trained neural network and allow a fine-tuning for the systemic parameters. Moreover, in \ref{fig7}(c), we also further demonstrate the effectiveness of detecting the strong phonon blockade effect when all systemic parameters have a small disturbance. The results show that the machine learning technique is suitable and effective for detecting the strong phonon blockade effect in this nonlinearly coupled system.

In the above description, we have proposed a scheme to observe the phonon blockade effect in a nonlinear coupled optomechanical system. Here we mainly discuss the experimental feasibility. The quadratically coupled optomechanical system has been studied in various platform and the mechanical characters of these platforms are usually observed by indirectly measuring the optical mode~\cite{PhysRevA.89.023849,PhysRevA.85.053832,PhysRevX.1.021011,PhysRevA.95.033803}. Generally, the small coupling coefficient needs to be enhanced by exploiting some experimental techniques.  However, the effective coupling strength $J=g\alpha$ is adjustable by controlling the flexible and strong driving field acting on the cavity mode. The quadratic coupling strength $g\simeq 245$~Hz has been experimentally realized in a planar silicon photonic crystal cavity \cite{PhysRevX.5.041024} and the coupling strength $g$ can arrive the strength from 1~kHz~\cite{Kalaee2016} to 100~kHz~\cite{PhysRevX.5.041024} by properly tuning of the double-slotted photonic crystal structure. Due to the large increase of $g$ in the photonic crystal cavity, the effective coupling strength $J$ can arrive 10 MHz to 100 MHz by controlling the amplitude of the strong optical driving field. The decay rates are $\gamma/2\pi=125$~kHz and $\kappa/2\pi=20$~MHz, respectively \cite{PhysRevX.5.041024}. The small thermal average phonon numbers $n_{th}\simeq 0.00001$ is also obtained in experiment with temperature 25~mK  and mechanical frequency 6~GHz\cite{OConnell2010}. Moreover, the decay of the optical cavity $\kappa/2\pi=0.66$~MHz and mechanical decay $\gamma/2\pi=1000$~Hz have been achieved in experiment~\cite{Murch2008}. Therefore, the condition for realizing the phonon blockade effect is possible to be achieved with current technology in a photonic crystal optomechanical cavity.

We proposed a scheme to detect phonon blockade in a quadratically coupled optomechanical system. The input of the feedforward neural network is from the homodyne detection, and the corresponding output is the equal-time second-order correlation function. Our training results show that the MSEs of training, testing, and validation sets monotonically decrease to a very small value, which indicates the trained neural network exhibit excellent performance in predicting the phonon blockade. Moreover, we also take some examples to evaluate the performance of the neural network. Our results show that the predicted values are approximately coincident with the ideal values under wide parameter scope. Therefore, our scheme effectively solves the challenge of detecting strong phonon blockade even the system exists a nonlinear interaction. Besides, our scheme is robust against the small disturbance of the system parameters. Therefore, our scheme has potential applications in building a phonon blockade detector in the future.

We thank Wenlin Li  for his fruitful discussions.  This work was support by National Natural Science Foundation of China (11574041, 11375036);  Natural Science Foundation of Liaoning Province (201801156).			


\bibliography{mybibtex}

\end{document}